\documentclass[aps,superscriptaddress,twocolumn,amsmath,amssymb]{revtex4-1}

\usepackage{graphicx}
\usepackage{epstopdf}
\usepackage{bm}
\usepackage{subfigure}
\usepackage{sidecap}
\usepackage{times}
\usepackage{graphicx}
\usepackage{epstopdf}
\usepackage{amssymb}
\usepackage{amsmath}

\usepackage{bm}
\usepackage{subfigure}
\usepackage{sidecap}
\graphicspath{{Figures/}}

\newcommand{\la}{\lambda}

\newcommand{\De}{\Delta}

\newcommand{\balp}{{\bm\alpha}}
\newcommand{\bbet}{{\bm\beta}}
\newcommand{\bsig}{{\bm\sigma}}

\newcommand{\bphi}{{\bm\phi}}

\newcommand{\bkap}{{\bm\kappa}}
\newcommand{\bchi}{{\bm\chi}}

\newcommand{\hbk}{\hat{{\bf k}}}

\newcommand{\be}{\begin{equation}}
\newcommand{\ee}{\end{equation}}

\newcommand{\bea}{\begin{eqnarray}}
\newcommand{\eea}{\end{eqnarray}}
\newcommand{\bd}{\begin{displaymath}}
\newcommand{\ed}{\end{displaymath}}
\newcommand{\ba}{\begin{array}}
\newcommand{\ea}{\end{array}}
\newcommand{\bi}{\begin{itemize}}
\newcommand{\ei}{\end{itemize}}
\newcommand{\bc}{\begin{center}}
\newcommand{\ec}{\end{center}}
\newcommand{\bfl}{\begin{flushleft}}
\newcommand{\efl}{\end{flushleft}}
\newcommand{\bfr}{\begin{flushright}}
\newcommand{\efr}{\end{flushright}}
\newcommand{\non}{\nonumber}

\newcommand{\bl}{\begin{aligned}}
\newcommand{\el}{\end{aligned}}

\newcommand{\hatt}{\hat{t}}
\newcommand{\hh}{\hat{h}}

\newcommand{\hE}{\hat{E}}

\newcommand{\tE}{\tilde{E}}

\newcommand{\tga}{\tilde{\gamma}}

\newcommand{\fs}{\frac{1}{2}}

\newcommand{\bse}{Bi$_2$Se$_3$}
\newcommand{\bte}{Bi$_2$Te$_3$}
\newcommand{\ste}{Sb$_2$Te$_3$}

\def\ket#1{\left\vert #1 \right\rangle}

\def\ba{{\bf a}}
 
\def\bk{{\bf k}}   
 
\def\bp{{\bf p}}

 \def\bd{{\bf d}}   
  \def\hbz{\hat{{\bf z}}}

\def\da{\downarrow} \def\ua{\uparrow}

\def\bra{\langle}
\def\ket{\rangle}
\def\={\!\!\!&=&\!\!\!}
\def\+{\!\!\!&&\!\!\!+~}
\def\-{\!\!\!&&\!\!\!-~}

\usepackage[colorlinks=true,citecolor=blue]{hyperref}

\usepackage{hyperref,cleveref}
\usepackage{color}
\usepackage[dvipsnames]{xcolor}
\usepackage{xcolor}

\begin{document}

\title{Surface step states and Majorana end states in profiled topological insulator thin films}

\author{Peter Thalmeier}
\affiliation{Max Planck Institute for the  Chemical Physics of Solids, D-01187 Dresden, Germany}

\author{Alireza Akbari}
\affiliation{Max Planck Institute for the  Chemical Physics of Solids, D-01187 Dresden, Germany}
\affiliation{Max Planck POSTECH Center for Complex Phase Materials, and Department of Physics, POSTECH, Pohang, Gyeongbuk 790-784, Korea}

\date{\today}

\begin{abstract}
The protected helical surface states in thin films of  topological insulators (TI) are subject to inter-surface hybridisation.
This leads to gap opening and spin texture changes as witnessed in photoemission and quasiparticle interference
investigations. Theoretical studies show that universally the hybridisation energy exhibits exponential decay as well
as sign oscillations as function of film thickness, depending on the effective band parameters of the material.
When a step is introduced in the TI film e.g. by profiling the substrate such that the hybridisation has different signs on 
both sides of the step, 1D bound states appear within the hybridisation gap which decay exponentially with distance
from the step. The step bound states have linear dispersion and inherit the helical spin locking from the surface
states and are therefore non-degenerate. When the substrate becomes an s-wave superconductor Majorana zero modes
located at the step ends are created inside the superconducting gap. The proposed scenario involves just
a suitably stepped interface of superconductor and  TI and therefore may be a most simple device being able
to host Majorana zero modes.

\end{abstract}

\maketitle

\section{Introduction}
\label{sect:intro}

The surfaces of strong topological insulators (TI) like \bse, \bte~and \ste~carry spin-locked non-degenerate helical surface 
states. As long as the surfaces are isolated, i.e. their distance is much larger than the surface state decay length
into the bulk the 2D dispersion is described by isotropic Dirac cones to lowest order in momentum counted from
the time-reversal invariant (TRI) points and by warped cones with sixfold symmetry due to higher order terms. They have been verified 
indirectly by magnetotransport measurements \cite{taskin:11,taskin:12} as well as directly
by photoemission \cite{hsieh:09,chen:09,kuroda:10,hoefer:14} and surface tunneling \cite{roushan:09,zhang:09,alpichshev:10,okada:11,cheng:12,kohsaka:17,lee:09} experiments, the latter were explained in numerous
theoretical investigations \cite{lee:09,zhou:09,ruessmann:20}.
This simple situation changes in an interesting manner when one considers TI thin films with thickness small enough so that
inter-surface hybridisation of bottom (B) and top (T) surface states occurs. Due to the interaction topological protection for
states close to the Dirac point is lifted and a hybridisation gap in the excitation spectrum opens. This has indeed been verified directly by photoemission \cite{zhang:10,neupane:14} but also by a sudden breakdown of weak anti-localisation in magnetotransport \cite{taskin:11,taskin:12} as function of film thickness when the latter falls below about five quintuple layers (QL). The effective hybridisation  between the surface states has been calculated \cite{lu:10,asmar:18,asmar:21} solving the thin film boundary value problem for an effective $\bk\cdot\bp$ Hamiltonian of the bulk. Due to the constrained film geometry
the effective inter-surface hybridisation $t(d)$ not only decreases exponentially with film thickness $d$ but also generally oscillates
as function of thickness d with an oscillation period depending on the material parameters. This should be again visible in the photoemission gap and in the concommitant oscillation of quasiparticle interference (QPI) patterns predicted in Ref. \onlinecite{thalmeier:20}.\\

%
\begin{figure}[t]
\begin{center}
\includegraphics[width=0.98\columnwidth]{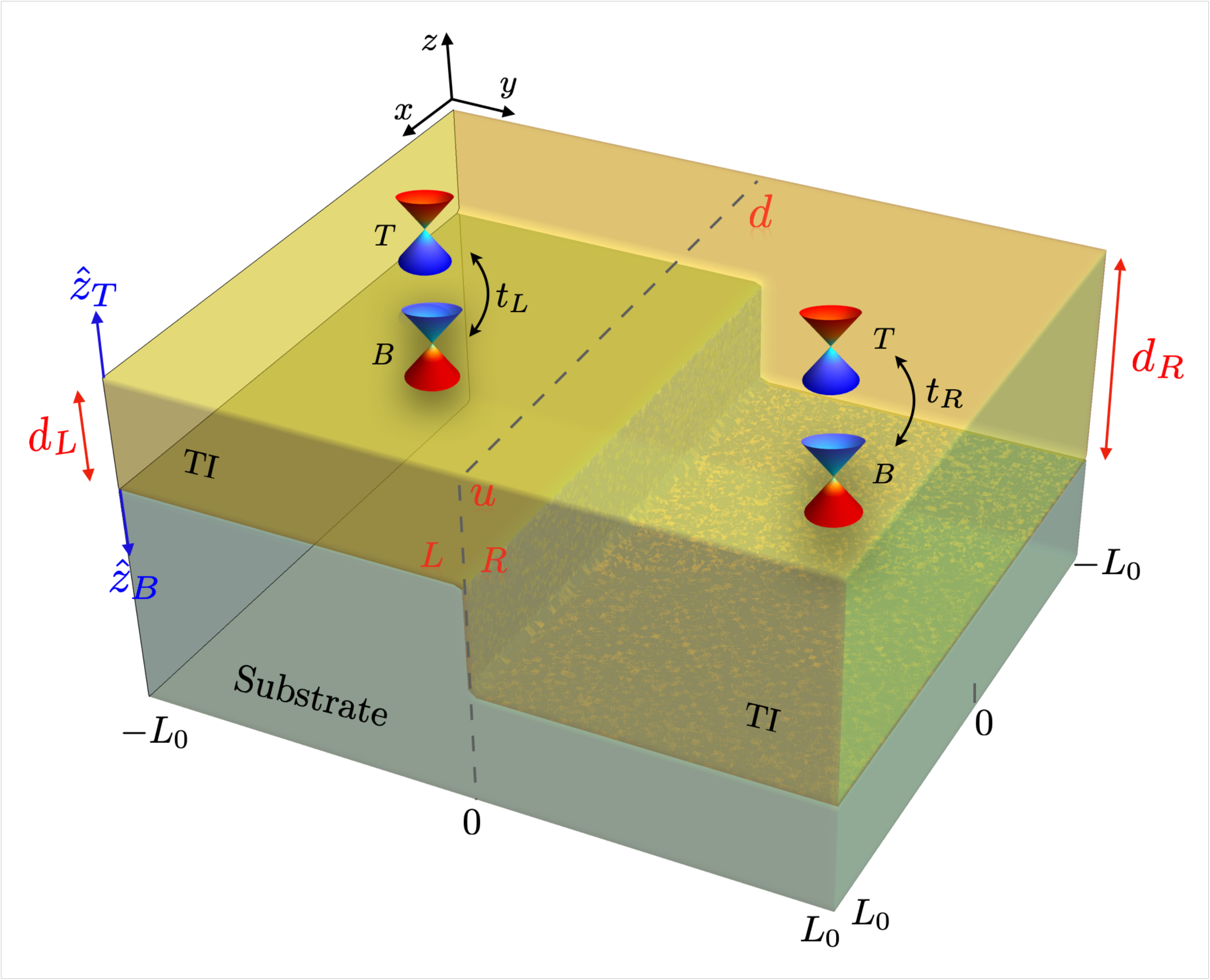}
\end{center}
\vspace{-0.5cm}
\caption{Schematic view of the TI step configuration due to profiled substrate, for simplicity equal x,y dimension $2L_0$ of the device is chosen. The TI film thicknesses $d_L$ and $d_R$ to the left and right of the step lead to different hybridisation strengths $t_L$ and $t_R$ between the Dirac cones on top (T) and bottom (B) surfaces of the TI film. For $t_L\cdot t_R <0$ this leads to the appearance
of a 1D bound state exponentially located at the step $y\approx0$ and extended along $x$. If the substrate is an s-wave superconductor Majorana zero modes at u,d positions may appear. For clarity film thickness and step size are exaggerated.
}
\label{fig:stepscheme}
\end{figure}
%

Aside from the modulated excitation gap at the Dirac point the oscillation of $t(d)$ gives rise to another novel and highly interesting scenario which is the subject of this investigation.
Suppose the film thickness is not constant but changes in a steplike manner at a certain lateral position (see Fig.~\ref{fig:stepscheme}). If the thickness to the left $(d_L)$ and 
right $(d_R)$ of the step is chosen in such a way that the hybridisation  $t_L(d_L)$ and $t_R(d_R)$ have {\it opposite} signs on the two sides of the step then a helical non-degenerate bound state within the hybridisation gap may appear which is spatially located at the step and has a linear dispersion. If found experimentally this would entail a further interesting speculative possibility: Once the substrate of the step-profiled TI becomes a  simple s-wave spin singlet superconductor the proximity effect will open a superconducting gap in the dispersion of nondegenerate (spin-locked) step state. This is a typical situation that can create Majorana end states. Such scenario have previously  investigated e.g. with TI nanowires on SC substrate \cite{sau:21,sau:10,stanescu:13,cook:11,cook:12,das:12} which may need the application of a magnetic flux through the wire. In alternative devices the wire has to be itself ferromagnetic \cite{livanas:19} or heterostructures with ferromagnetic layers are used \cite{stanescu:13}. In the present scenario no magnetic field has to be applied, a properly chosen step of the TI profile on the substrate is sufficient to  create the possibility for Majorana states at the step ends. This seems to be a most simple way to realise these states in a realistic geometry.

\section{Isotropic Dirac surface state model for TI} 
\label{sect:model}

As a starting point for the homogeneous thin film we use the isotropic TI surface state model on the 
top (T) and bottom (B) surface of the film. We will use $\alpha, \sigma, \kappa$ indices and associated 
Pauli matrix vectors $(\balp,\bsig,\bkap)$ to denote two dimensional T/B surface, $|\uparrow\rangle,|\downarrow\rangle$ spin and  $|\pm1\rangle$ 
helicity spaces. The unit in each space is denoted by $(\alpha_0,\sigma_0,\kappa_0)$.
On a single isolated surface, using the operator basis $\psi^{\dag}_\bk=(c^{\dag}_{\ua\bk},c^{\dag}_{\da\bk})$, the isotropic 2D model Hamiltonian is given by
\be
\bl
\!
{\cal H}=
\!
\sum_\bk \psi^\dag_\bk h_\bk\psi_\bk;\;\;
h_\bk=v(\bk\times\bsig)\cdot\hbz =
v\left(
 \begin{array}{cc}
0& -ik_-\\
ik_+& 0
\end{array}
\right),
\label{eqn:hamat}
\el
\ee
where $\bk=(k_x,k_y)$ is the wave vector counted from the TRI Dirac point, $\bsig$ the spin, $\hbz$ the surface normal and
$v$ the velocity, furthermore $k_\pm=k_x\pm i k_y$ and $k=|\bk|=(k_x^2+k_y^2)^\fs$. This expression is proportional to
the helicity operator $\kappa_{\hbk}=(\bsig\times\bk)\cdot\hbz/k$, namely $h_\bk=(vk)\kappa_{\hbk}$.
The eigenvalues or dispersions of
the Dirac cone and associated eigenvectors in the spin basis $|\uparrow\rangle,|\downarrow\rangle$ are described by
\bea
\epsilon^\pm_\bk=\pm(vk);\;\;\;
S_\bk=\frac{1}{\sqrt{2}}
\left(
 \begin{array}{cc}
1& ie^{-i\theta_\bk}\\
 -ie^{i\theta_\bk}& 1
\end{array}
\right),
\eea
where the columns of the unitary matrix $S_\bk$ are the helical eigenstates $|\pm1\rangle$. In this basis the helicity operator 
is simply represented by the Pauli matrix $\kappa_{\hbk}=\kappa_z$. Furthermore $\theta_\bk=\tan^{-1}(k_y/k_x)$ is the azimuthal angle of the \bk-vector.

\section{Hybridisation and gap opening of surface states in TI thin films}
\label{sect:film}

In a TI film with thickness $d$ much larger than the surface state decay length the top and bottom surfaces
may be considered as independent surface states. One only has to keep in mind that surface normals are
oppositely oriented to the global $\hbz$ direction, i.e. $\hbz_T=-\hbz_B \equiv\hbz$. Therefore helicities for both
energies are also opposite on T, B surfaces according to $\kappa^B_{\hbk}=-\kappa^T_{\hbk}$ and
 therefore $h^B_\bk=-h^T_\bk\equiv- h_\bk$. 
 However, when the film thickness is reduced (below a few quintuple layers) the T,B surface states overlap and
 a hybridisation of equal spin (and therefore equal helicity) eigenstates develops. This problem has been fundamentally treated
 and analyzed in great generality in the work of Asmar et al \cite{asmar:18} and also in Ref.~\onlinecite{shan:10}. Here we use a simplified model with a \bk-independent effective hybridisation element $t(d)$ that depends, however, on film thickness $d$. The thin film surface states Hamiltonian is
 then given in spin representation~ \cite{thalmeier:20} using  $\Psi^{\dag}_\bk=(c^{T\dag}_{\ua\bk},c^{T\dag}_{\da\bk},c^{B\dag}_{\ua\bk},c^{B\dag}_{\da\bk})$,
 \be
 \bl
 {\cal H}
 &
 =\sum_\bk \Psi^\dag_\bk \hat{h}_\bk\Psi_\bk;\;\;
 \\
 \hat{h}_\bk
 &
 =
 v(k_x\sigma_y-k_y\sigma_x)\alpha_z+t\sigma_0\alpha_x
\\
&
 =
 \left(
 \begin{array}{cc}
v(k_x\sigma_y-k_y\sigma_x)& t\sigma_0\\
 t \sigma_0& -v(k_x\sigma_y-k_y\sigma_x)
 \end{array}
\right),
\el
\label{eqn:hamfilm}
 \ee
 or in helicity representation according to
 \be
 \bl
 \hat{h}_\bk=\epsilon_\bk\kappa_z\alpha_z+t\kappa_0\alpha_x=
 \left(
 \begin{array}{cc}
\epsilon_\bk\kappa_z& t\kappa_0\\
 t \kappa_0& -\epsilon_\bk\kappa_z
 \end{array}
\right),
 \el
 \ee
 The eigenvalues of the film Hamiltonian are then obtained as
\bea
E^\pm_\bk=\pm[(vk)^2+t(d)^2]^\fs
.
\label{eqn:enhyb}
\eea
Which exhibit a thickness dependent hybridisation gap $t(d)$ at the Dirac point $\bk=0$. Each of these
dispersion branches is twofold degenerate which is inherited from the (T,B) degeneracy of states in
the uncoupled $(t=0)$ case.
The degeneracy is lifted if the bottom surface experiences an effective bias due to the substrate effect. This 
can be described by adding a term $\Delta_{su}\sigma_0\alpha_z$ to Eq.~(\ref{eqn:hamfilm}). Then the
split hybridised surface bands are given by
\be
\bl
E^\pm_{\bk 1,2}=\pm[(|vk|\pm|\Delta_{su}|)^2+t(d)^2]^\fs,
\el
\ee
where split band indices $1,2$ refer to $\pm$ inside the square root. Since the above substrate term can in principle 
be canceled by an applied bias voltage $({\rm eV})$ term at the substrate  we will keep the degenerate thin film model of Eq.~(\ref{eqn:enhyb}). Furthermore the chemical potential $\mu$ can be controlled by applying a gate voltage at the TI surface.
\subsection{The oscillation model of hybridisation with film thickness}
\label{subsect:hybosc}

\begin{figure}[t]
\includegraphics[width=0.90\columnwidth]{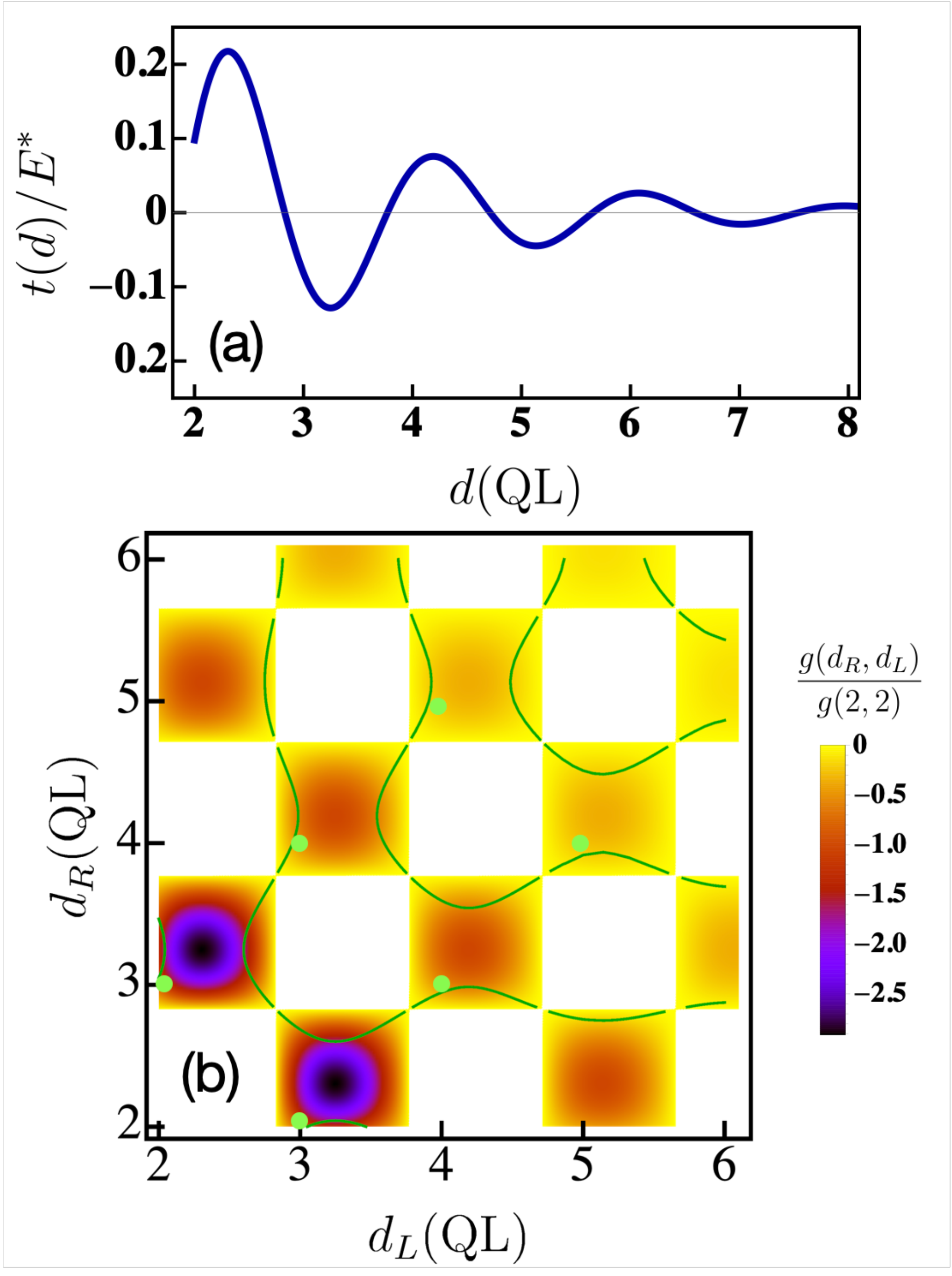}
\caption{(a) Film thickness dependence of inter-surface hybridisation energy in \bte~for the theoretical model parameters 
$(t_0,d_0,d'_0)=(0.8,1.79,0.3)$ corresponding to  Ref.~\onlinecite{asmar:18}  (in units of $E^*=0.25$ eV for energy and QL for thicknesses \cite{thalmeier:20}). (b) Contour plot of the function
$g(d_L,d_R)$ in Eq.~(\ref{eqn:existence}). In the coloured region a step bound state exists while it is absent in the white
regions. The darker color correspond to more tightly bound wave functions in Fig.~\ref{fig:envelope}. The contour lines denote
pairs $(d_R,d_L)$ with $t_R=-t_L$ leading to a symmetric step bound state around $y=0$. The green dots designate  pairs with integer thickness [QL] close to this line of symmetry.
}
\label{fig:hybridisation}
\end{figure}

The thickness dependent effective hybridisation $t(d)$ may be obtained from the solution of a subtle
boundary value problem for the thin film \cite{asmar:18}, starting form the $\bk\cdot\bp$ Hamiltonian of the bulk bands.
It may be represented by the phenomenological form \cite{asmar:18,thalmeier:20}
\be
\bl
t(d)=t_0\exp\bigl(-\frac{d}{d_0}\bigr)\sin\bigl(\frac{d}{d'_0}\bigr).
\label{eqn:tosc}
\el
\ee
Here the energy $t_0$ and thickness $d_0,d'_0$ scales are determined by the parameters of bulk bands \cite{asmar:18,thalmeier:20}.
As an example we give the theoretical values  for \bte~in terms of natural units $E^*=0.25$ meV and $1\mbox{QL}=10.16$\AA, respectively as $(t_0,d_0,d'_0)=(0.80,1.79,0.3)$. As expected the expression contains an exponential decay with increasing thickness $d$ but, in order to satisfy boundary conditions, also an oscillatory term, whose physical origin was derived in Ref.~\cite{asmar:18}.  The hybridisation is obtained as a perturbation integral of an effective inter-surface tunneling Hamiltonian connecting the uncoupled surface state wave functions. The decay length of the latter perpendicular to the surface may become a complex number, depending on the bulk band parameters. This leads to oscillating decay of the wave function which is inherited
by the hybridisation integral. We note that the vanishing t(d) for thin films at the nodes (Fig.~\ref{fig:hybridisation}) does not mean the surface states should be considered as decoupled as for large d in the bulk case  but they rather indicate the destructive interference of both surface wave function in the hybridisation integral.
These thickness oscillations of $t(d)$  play an essential role in the present investigation and its consequences have before been studied in view of its influence on quasiparticle interference (QPI) patterns \cite{thalmeier:20}. According to the theoretical estimation of parameters \cite{asmar:18,asmar:21} the oscillations of the gap $|t(d)|$ should be pronounced in \bte~and \ste~but not in \bse. 
 In the latter only half an oscillation period appears which is strongly damped by the rapid exponentioal decay of $t(d)$. The observable quantity is the (rectified) oscillation of inter-surface hybridisation gap $|t(d)|$. Since it can reasonably only be compared for films with identical surface terminations one is restricted to a discrete set of values for $t(d)$ with integer multiples of quintuple layers. Therefore weak oscillations of $|t(d)|$ may not easily be identified.
 The hybridisation gap opening as function of film thickness with integer number of QL  has been investigated by ARPES \cite{zhang:10} for \bse. The predicted exponential decay of $|t(d)|$ with increasing film thickness was observed but no clear evidence for the half oscillation period was found.
It was argued \cite{thalmeier:20} that a modest change of the theoretical bulk band parameters leading to different values of $d_0, d'_0$ in Eq.~(\ref{eqn:tosc}) could account for the suppression of the half oscillation.  
In fact charge transfer and bulk band bending effects due to the substrate have not been included in the theoretical model employed here but may influence the surface state energies \cite{zhang:10} and possibly the above oscillation parameters.
However, the origin of hybridisation gap oscillation is of universal nature enforced by boundary conditions due to the constrained geometry of thin films \cite{asmar:18}  and therefore it may appear whenever the bulk band parameters of the 3D TI lead to suitable $d_0, d'_0$ scales  with $d_0\gg d'_0$, i.e. a considerable number of oscillations before $t(d)$ decays. Therefore they should appear in TI material with more favourable scale parameters such as is prediced for  \bte~or \ste~(Fig.~\ref{fig:hybridisation}a). Sofar no systematic film thickness variation of the hybridsation gap in these materials has been investigated with either ARPES or QPI methods.

Due to their universal origin we are confident that the gap oscillations will eventually be found in thin films of a suitable TI material.
This expectation is the starting point for the following investigation of intriguing appearance of 1D topological states for profiled thin film geometry.

%
%

\section{Creation of 1D step bound states inside the TI thin film gap}
\label{sect:1Dbound}

The TI surface state form linearly dispersing 2D bands or Dirac cones inside the bulk gap $\Delta_b$ (Fig.\ref{fig:gapping}a) of the 
3D TI material which is due to bulk spin-orbit coupling. For sufficiently thin films the Dirac cones themselves
are opening a gap given by the size of the inter-surface hybridisation (Fig.\ref{fig:gapping}a). One might ask whether
the creation of topologically protected in-gap states can be repeated by some means in a `Matrjoschka'-like fashion, creating 1D linearly
dispersing  bands within the hybridisation gap of 2D surface states. One way is to create domain walls of some sort on the surface with
suitable properties. The easiest way to achieve a domain wall  is by stepping the thin film (e.g. by stepping the substrate) so that the film thickness and hybridisation are different on both sides of the step (see illustration in Fig.~\ref{fig:stepscheme}). This  possibility will be investigated in the following sections. 

\subsection{Boundary and existence conditions and dispersion of bound states}
\label{subsect:1Ddisp}

\begin{figure}[t]
\begin{center}
\includegraphics[width=\columnwidth]{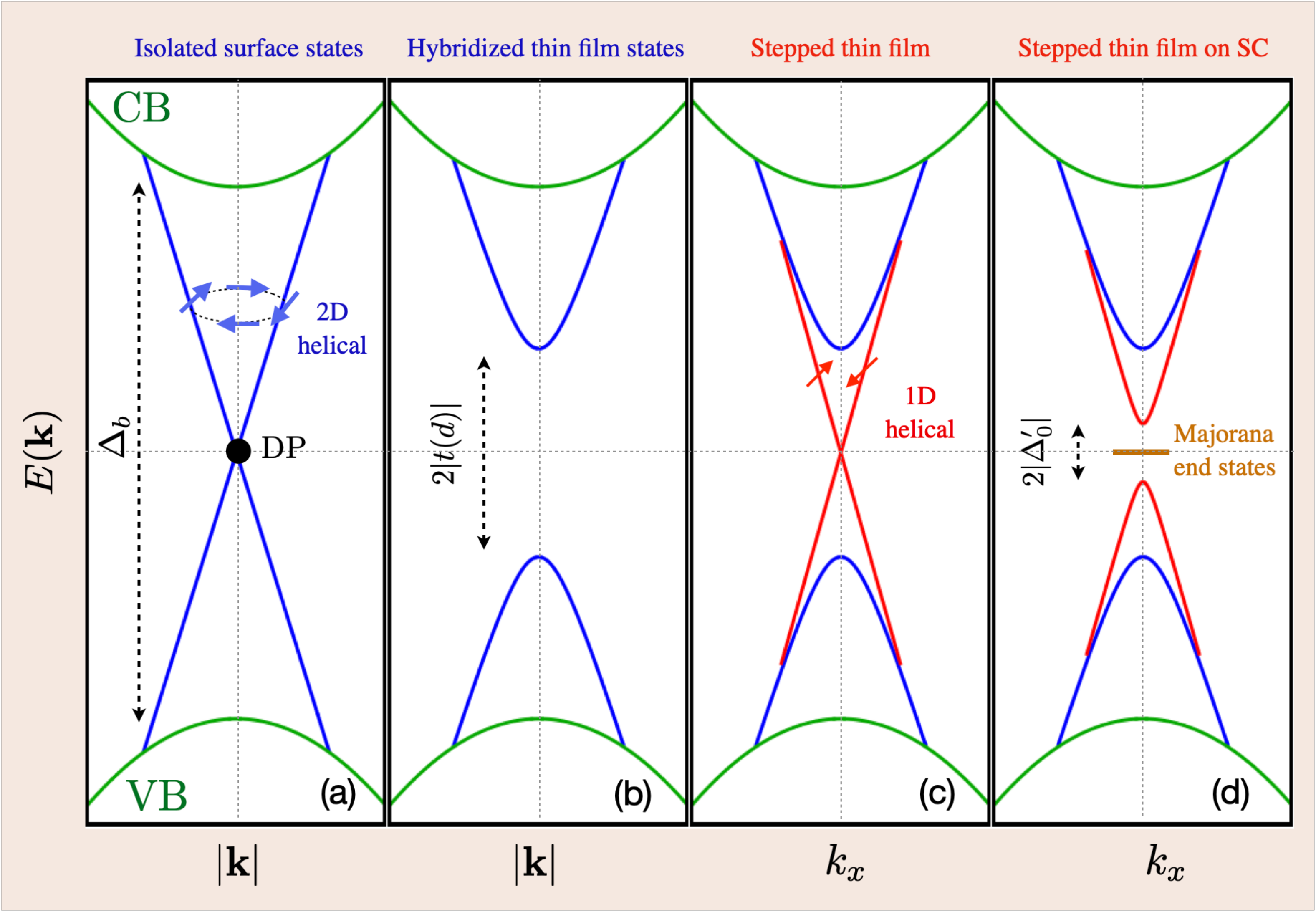}
\end{center}
\vspace{-0.5cm}
\caption{Schematic sequence of 2D surface state and 1D step state gappings. (a) ungapped 2D helical states on isolated surfaces.
(b) Gapping in homogeneous thin film due to inter-surface hybridisation $t(d)$. (c) Appearance of 1D helical step states (red) in profiled TI thin film  with velocity $v$ asymptotically equal to that of 2D hybridized surface states (blue). (d) Opening of SC proximity gap in the 1D step states and appearance of zero energy Majorana end states. Here $\Delta_b$ is the overall 3D bulk gap ( $\Delta_b/E^*\simeq 1$ for \bte). Note that for clarity the various gaps $\Delta_b, 2|t(d)|, 2|\Delta'_0|$ are not drawn to scale.
}
\label{fig:gapping}
\end{figure}

Let us adopt a  step geometry with the step extending along the $x$-direction, separating the left (L) and right (R) regions of the surfaces at $y=0$ and the surface normals oriented parallel to $z$  (Fig.~\ref{fig:stepscheme}). Then, assuming a sharp step the $y$-dependent inter-surface hybridisation is given by
\be
\bl
t(y)=t_L\Theta(-y)+t_R\Theta(y),
\label{eqn:sharps}
\el
\ee
where $t_L=t(d_L)$ and  $t_R=t(d_R)$, we assume $2L_0$ is the sample length along $x,y$  (Fig.~\ref{fig:stepscheme}). At the moment we keep the relative size of $d_L,d_R$ arbitrary. To find out whether a localised state at
the step develops one has to replace $k_y\rightarrow (-i\partial_y)$ and $t\rightarrow t(y)$ in the  Hamiltonian leading to the effective 1D problem 
(with $k_x$ parallel to the step treated as a fixed parameter) described in spin-surface space by
 \be
 \bl
 \hat{h}(k_x,y)
 =&
 v(k_x\sigma_y+i\sigma_x\partial_y)\alpha_z+t(y)\sigma_0\alpha_x
 \\
 =&
 \left(
 \begin{array}{cc}
v(k_x\sigma_y+i\sigma_x\partial_y)& t(y)\sigma_0\\
 t(y) \sigma_0& -v(k_x\sigma_y+i\sigma_x\partial_y)
 \end{array}
\right),
\label{eqn:hamstep}
\el
 \ee
If a localised bound state exists at the step it has to fulfil the envelope equation
\be
\bl
\hh(k_x,y)\bphi(k_x,y)=E\bphi(k_x,y)
\label{eqn:envelope}
\el
\ee
with $|E|<|t_L|,|t_R|$, i.e., lying inside the hybridisation gap of 2D surface states.
It is convenient to introduce rescaled energies $\hE=E/v$ and $\hatt=t/v$ which have the dimension
of wave number or inverse length (units $\mbox{QL}^{-1}$). The wave function is a four spinor defined by ($tr$ for transposed):
\be
\bl
\bphi^{tr}(k_x,y)=
\Big(
\phi^T_\ua(k_x,y),\phi^T_\da(k_x,y),\phi^B_\ua(k_x,y),\phi^B_\da(k_x,y)
\Big).
\el
\ee
For the the L,R sides of the step we use the following  ansatz to solve Eq.~(\ref{eqn:envelope}) $(\lambda=L,R)$:
\be
\bl
\phi_\lambda(k_x,y)=\ba_\lambda e^{ik_xx}e^{ik_y^\lambda y};\;\; 
\ba_\lambda^{tr}=(a^{T\lambda}_\ua,a^{T\lambda}_\da,a^{B\lambda}_\ua,a^{B\lambda}_\da),
\label{eqn:wavedef}
\el
\ee
where $k_y^L=-i\kappa_L$ and $k_y^R=i\kappa_R$ are given by the inverse decay lengths $\kappa_R, \kappa_L$ of the bound state localised at the step. Inserting into Eq.~(\ref{eqn:envelope}) we obtain
\be
\bl
\hE^2=k_x^2-\kappa_\la^2+\hatt_\la^2;\;\;\; \kappa^2_L-\kappa^2_R=\hatt^2_L-\hatt^2_R
,
\label{eqn:energy}
\el
\ee
The second equation follows because the first one has to be fulfilled for both sides L and R simultaneously.
The remaining relation to determine $\kappa_\alpha$ is obtained from the boundary condition at the step
according to $\phi_L(k_x,0)=\phi_R(k_x,0)$.
  These $\lambda= L,R$ wave functions for step states with energies $\hE$ (Eq.~(\ref{eqn:energy})) are determined by the 
solutions of
\begin{widetext}
\bea
\left(
 \begin{array}{cccc}
-\hE& i(\kappa_\lambda-k_x)&\hatt_\lambda &0\\
 i(\kappa_\lambda+k_x) & -\hE &0&\hatt_\lambda\\
 \hatt_\lambda&0&-\hE& -i(\kappa_\lambda-k_x)\\
 0&\hatt_\lambda& -i(\kappa_\lambda+k_x)&-\hE
 \end{array}
\right)
 \left(
 \begin{array}{c}
a_\ua^{T\lambda}\\
a_\da^{T\lambda}\\
a_\ua^{B\lambda}\\
a_\da^{B\lambda}
 \end{array}
\right)=0,
\eea
\end{widetext}
The matrix has rank 2 and therefore 2 components $a_1\equiv a_\ua^{T\lambda}$ and $a_2\equiv a_\da^{T\lambda}$ 
may be considered as free parameters 
for the solution. The other two components are obtained as 
\be
\bl
a^\lambda_{B\ua}
=&
\frac{\hE}{\hatt_\lambda}a_1-\frac{i(\kappa_\lambda-k_x)}{\hatt_\lambda}a_2;\;\;\;
\\
a^\lambda_{B\da}
=&
\frac{\hE}{\hatt_\lambda}a_2-\frac{i(\kappa_\lambda+k_x)}{\hatt_\lambda}a_1.
\label{eqn:amprelation}
\el
\ee
The  ratio $a_1/a_2$ is fixed  by the continuity condition  $\phi_L(k_x,0)=\phi_R(k_x,0)$. From the two equations above we
obtain
\bea
\bl
\frac{a_1}{a_2}
=&i\frac{\hatt_L(\kappa_R+k_x)+\hatt_R(\kappa_L-k_x)}{\hE(\hatt_R-\hatt_L)};
\\
\frac{a_2}{a_1}
=&i\frac{\hatt_L(\kappa_R-k_x)+\hatt_R(\kappa_L+k_x)}{\hE(\hatt_R-\hatt_L)}.
\label{eqn:ampratio}
\el
\eea
Taking the product and using the symmetrised Eq.~(\ref{eqn:energy}) with $\hE^2=k_x^2-\fs[(\kappa_R^2+\kappa_L^2)-(\hatt^2_R+\hatt^2_L)]$,
 we finally arrive at the second relation
\be
\bl
\fs\bigl[(\kappa_R^2+\kappa_L^2)-(\hatt_R^2+\hatt_L^2)\bigr]=
\Bigl(\frac{\hatt_L\kappa_R+\hatt_R\kappa_L}{\hatt_R-\hatt_L}\Bigr)^2
.
\label{eqn:bcond}
\el
\ee
This equation together with the one in Eq.~(\ref{eqn:energy}) may be solved by
expressing them in terms of the (anti-) symmetrised quantities $\fs(\kappa_R\pm\kappa_L)$. 
After some algebra one obtains the simple relations
\be
\bl
\kappa_R
=&
\fs|\hatt_R-\hatt_L|+\fs(\hatt_R+\hatt_L)sign(\hatt_R-\hatt_L);
\\
\kappa_L
=&
\fs|\hatt_R-\hatt_L|-\fs(\hatt_R+\hatt_L)sign(\hatt_R-\hatt_L).
\el
\ee
For a localised step state within the thin film hybridisation gap one must have both $\kappa_R,\kappa_L>0$. It is easy
to see from the above expressions that this can only  be possible if the we have the fundamental relation
\begin{widetext}
\be
\bl
g(d_R,d_L)=\hatt_R(d_R)\cdot\hatt_L(d_L)<0 :\;\;\;\;
\left\{
\begin{array}{rl}
\hatt_L<0<\hatt_R \;\;(\tau_{RL}=+1);   \;\;\;&    \kappa_R=\hatt_R;\;\;\;  \kappa_L=-\hatt_L=|\hatt_L|\\
\hatt_R<0<\hatt_L \;\;(\tau_{RL}=-1);   \;\;\;&    \kappa_L=\hatt_L;\;\;\;  \kappa_R=-\hatt_R=|\hatt_R|
\label{eqn:existence}
\end{array}
\right.
\el
,
\ee
\end{widetext}
where we defined $\tau_{RL}=sign(\hatt_R-\hatt_L)$. Inserting $\kappa_{R,L}$ for these to cases
into  Eq.~(\ref{eqn:energy}) for the step state energy we simply get, after reordering, the two linear dispersing
energies:
\be
\bl
E_{k_x\pm}=\pm(vk_x),
\label{eqn:E1D}
\el
\ee
which form two linear dispersing branches of 1D excitations localised at and moving along the step with
an energy inside the thin-film hybridisation gaps for $|k_x|<\min\{|\hatt_R|,|\hatt_L|\}$. 
We note that
the dispersion relation, i.e. the velocity $v$ is {\it independent} of the size of hybridisations $t_L,t_R$ and hence of the asocciated asymmetric decay of the wave function perpendicular to the step.
This non-degenerate step state as enforced by the boundary conditions exists as long as $\hatt(y)$ {\it changes sign} when crossing
the step. It is therefore topologically protected as long as this condition is fulfilled. These 1D topological states
inside the gap of hybridised 2D topological thin film states are schematically shown in Fig.~\ref{fig:gapping}(c).
 The velocity $v$ of 1D excitations (red) is asymptotically the same
as those of the gapped and hybridized 2D surface excitations (blue).
%
%

\subsection{Eigenvectors and helicity of the step states}
\label{subsect:helicity}

The four amplitudes  $\ba_\lambda$ of the wave function in Eq.~(\ref{eqn:wavedef}) are obtained from 
Eqs.~(\ref{eqn:amprelation},\ref{eqn:ampratio}) and the condition that $\langle \ba_\lambda^\dag |\ba_\lambda\rangle=1$. They are
the same on both sides L,R. For the two orthogonal states corresponding to $\hE_\pm(k_x)=\pm k_x$, we obtain
\bea
\ba_+=\fs
 \left(
 \begin{array}{c}
1\\
i\\
-\tau_{RL}\\
i\tau_{RL}
 \end{array}
\right);\;\;\;
\ba_-=\fs
 \left(
 \begin{array}{c}
1\\
-i\\
\tau_{RL}\\
i\tau_{RL}
 \end{array}
\right)
.
\label{eqn:1Dvector}
\eea
The complete normalised wave functions of the two nondegenerate step states are finally given by
\be
\phi_\pm(k_x,y)=\frac{1}{\nu_0}\ba_\pm[e^{\kappa_Ly}\Theta(-y)+ e^{-\kappa_Ry}\Theta(y)]e^{ik_xx},
\label{eqn:stepstate}
\ee
where the normalisation is  $\nu_0=L_0^\fs(\kappa^{-1}_L+\kappa^{-1}_R)$
with $L_0$ denoting half the step length in $x$-direction and $\kappa_{L,R}$ corresponding to the two possible
cases of Eq.~(\ref{eqn:existence}). Examples of the step wave functions for various integer QL thickness pairs 
$(d_R,d_L)$ are given in Fig.~\ref{fig:envelope}(a).
These wave functions are eigenstates to the helicity operator.
Since for the step states $\bk=k_x\hat{{\bf x}}$ the latter is given by $\kappa_z=-\sigma_y\fs(1+\alpha_z)+\sigma_y\fs(1-\alpha_z)\equiv -\sigma_y^T+\sigma_y^B$.
The spinors $\ba_\pm$ (and also the total wave functions $\phi_\pm$) then fulfil $\kappa_z\ba_\pm=\mp\ba_\pm$.
This property is inherited from the isolated T,B film states. Because of the 1D character of step states it means the spin is always
locked perpendicular to $\bk=k_x\hat{{\bf x}}$, i.e. parallel to $y$. It is also  useful to consider the expectation values of the spin $\bsig$.
One finds $\bra \ba_\pm|\sigma_y^T|\ba_\pm\ket=\pm\fs$ and  $\bra \ba_\pm|\sigma_y^B|\ba_\pm\ket=\mp\fs$. All other spin expectation values vanish. The eigenvectors $\ba_\pm$ define the field operators $\Psi_\lambda(x,y)$ $(\lambda=\pm)$ of helical step states according to (now using $k=k_x$ for 1D states):
\be
\bl
\Psi_\lambda(x,y)
=&
\sum_k \phi_\la^\dag(k,y)\Psi_k
\\
=&
\sum_k\chi_{k\la}\frac{1}{\nu_0}
[e^{\kappa_Ly}\Theta(-y)+ e^{-\kappa_Ry}\Theta(y)]e^{ikx},
\el
\ee
where the quasiparticle operator algebra for the 1D helical step states is  defined by $\chi_{k\la}=\ba^\dag_\la\Psi_k$ which is explicitly
given by (using the abbreviation $\tau=\tau_{RL}=sign(\hatt_R-\hatt_L)$ in Eq.~(\ref{eqn:1Dvector}));
\be
\bl
\chi_{k+}
=&\fs(c^T_{k\ua}-ic^T_{k\da}-\tau c^B_{k\ua}-i\tau c^B_{k\da});
\\
\chi_{k-}
=&\fs(c^T_{k\ua}+ic^T_{k\da}+\tau c^B_{k\ua}-i\tau c^B_{k\da}).
\label{eqn:1Dquasi}
\el
\ee
They fulfil the canonical anti-commutation relations $\{\chi_{k\la},\chi_{k'\la'}^\dag\}=\delta_{kk'}\delta_{\la\la'}$. In terms of these 1D quasiparticle operators the 1D step state Hamiltonian may be written as (cf. Eq.(\ref{eqn:E1D}))
\be
H_{ST}=\sum_{k\la}E_{k\la}\chi_{k\la}^\dag\chi_{k\la}.
\label{eqn:1DHam}
\ee
These step states form the basis to construct Majorana end states through the proximity effect originating from the superconducting substrate. Before this, however we consider the situation for a more realistic step profile.

\subsection{Extension to bound states for soft steps}
\label{subsect:softstep}

\begin{figure}[t]
\begin{center}
\includegraphics[width=0.95\columnwidth]{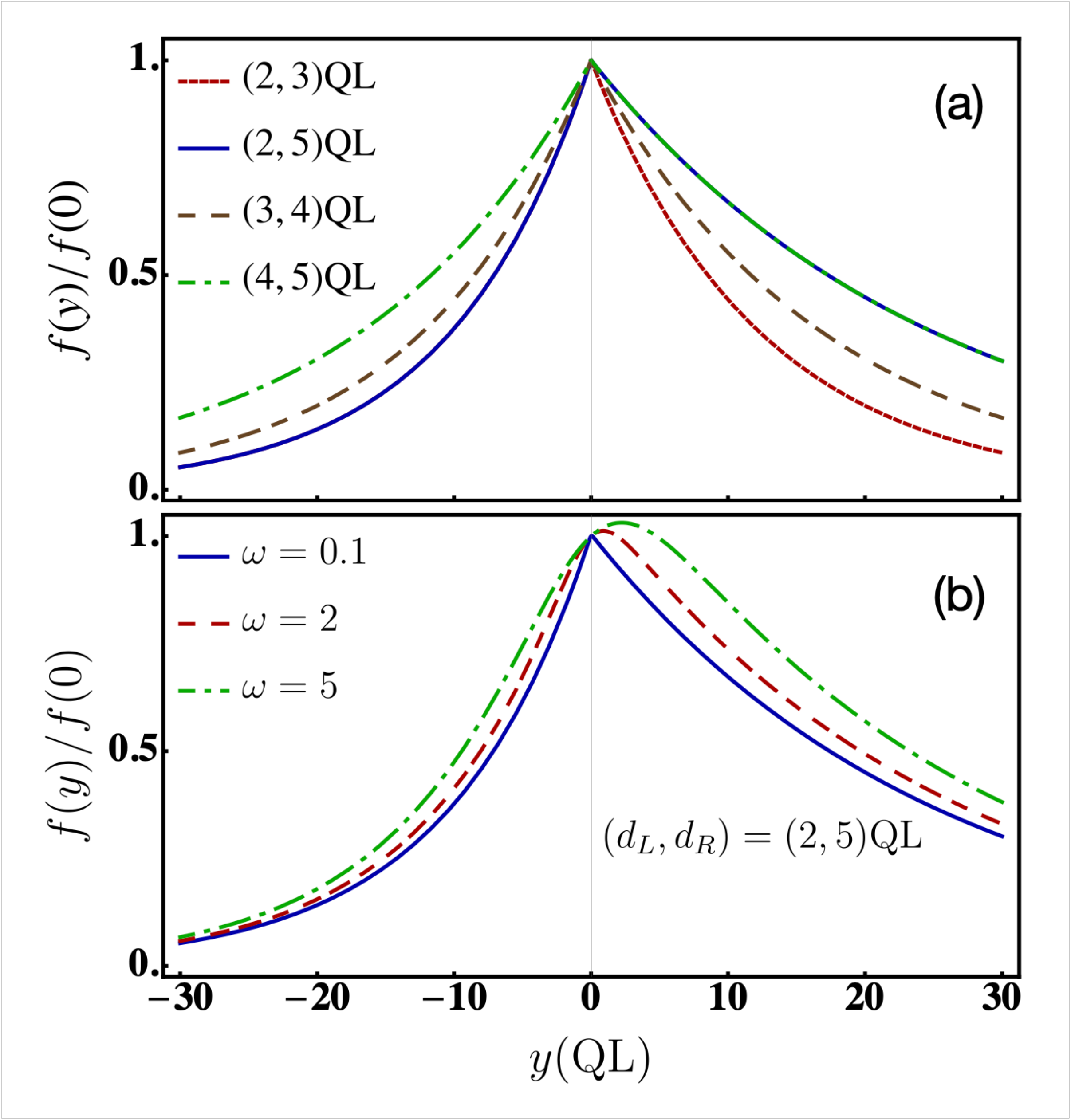}
\end{center}
\vspace{-0.5cm}
\caption{Step state envelope function $f(y)$ (Eq.~(\ref{eqn:softenv})). (a) for different thicknesses $d_R,d_L$ and a sharp 
step ($w=0$). In this case the ordinate is also equal to $|\phi_\pm(k_x,y)|/|\phi_\pm(k_x,0)|$ (Eq.~\ref{eqn:stepstate})).  The 
pair $(d_R,d_L)=(2,3)$ is not far from the symmetric step state. (b) fixed thickness pair $(2,5)$ but different step 
widths tuned by $w$ [QL].
}
\label{fig:envelope}
\end{figure}

The existence of the 1D step states is not tied to having a sharp step. A more softer profile serves the same purpose,
for example replacing Eq.~(\ref{eqn:sharps}) with a soft step of width $w$:
\bea
\label{eqn:softstep}
\bl
\non
\hatt(y)=\fs(\hatt_R+\hatt_L)+\fs(\hatt_R-\hatt_L)\tanh\frac{y}{w}
\left\{
\begin{array}{rl}
&\hatt_L; \;\;\; y\gg w \\
&\hatt_R; \;\;\; y\ll w
\label{eqn:softs}
\end{array}
\right.
\el
\\
\eea
To solve the wave equation Eq.~(\ref{eqn:envelope}) we now make a smooth envelope function ansatz instead of separating 
between L,R regime (Eq.~(\ref{eqn:wavedef})). It may be written as 
\be
\phi_\lambda(k_x,y)=\frac{1}{\nu}\ba_\lambda e^{ik_xx}f(y);
\label{eqn:softwave}
\ee
with
\be
\bl
f(y)
=
&
\exp
\Big[
-\tau_{RL}\int_0^y\hatt(y')dy'
\Big]
\\=&
\exp
\Big[
-\tau_{RL}
\Big(
\bar{\hatt}y+\hatt'w\ln\cosh\frac{y}{w}
\Big)
\Big].
\label{eqn:softenv}
\el
\ee
The form of spinors in Eq.(\ref{eqn:softenv}) and the dispersions is the same as for the model with a sharp step.
Here we defined the symmetrised expressions $\bar{\hatt}=\fs(\hatt_R+\hatt_L)$ and  $\hatt'=\fs(\hatt_R-\hatt_L)$. 
A comparison of step and envelope functions for various widths $w$ is presented in Fig.~\ref{fig:envelope}(b).

\subsection{Influence of the 2D warping term}
\label{subsect:warping}

Our investigation of step states is based on the underlying assumption that the influence of higher order warping terms
for the 2D surface states can be neglected in the 1D step state formation.  Firstly it is well known how to include them in the surface state formation of thin films where they modify the isotropic Dirac cones and circular Fermi surface into cones and Fermi surface with six-pronged `snowflake' shape. A complete theory for homogeneous thin films, including the lowest order Dirac term (Eq.~(\ref{eqn:hamat})), the warping term and the inter-surface hybridisation on the same footing has been developed in Ref.~\cite{thalmeier:20} leading to gapped and warped Dirac cones for the 2D thin film quasiparticles. The remaining question of importance here is how the warping will influence the formation of the 1D step states. We may understand this in a straightforward way by treating the warping as a perturbation (which vanishes for $k_x\rightarrow 0$). In the inhomogeneous film geometry the warping term in spin-surface presentation is given by
 \bea
 \bl
 \hat{h}^w(k_x,y)
 =&
 \lambda D_y(k_x)\sigma_z\alpha_0;
 \\
 D_y(k_x)=&k_x(k_x^2+3\partial_y^2),
 \el
 \eea
which has to be added to the unperturbed Hamiltonian in Eq.~(\ref{eqn:hamstep}). Here $\lambda$ is the warping parameter (for realistic values in TI see Ref.~\cite{thalmeier:20}). Then, using the 1D step wave functions of Eq.~(\ref{eqn:stepstate}) the correction to the 1D step state dispersion Eq.~(\ref{eqn:E1D}) is given
by 
\bea
\bl
\delta E_{k_x\pm}=\int dy \phi^\dag_\pm(k_x,y) \hat{h}^w(k_x,y) \phi_\pm(k_x,y).
\el
\eea
Writing the four component eigenvectors $\ba_\pm$ in Eq.~(\ref{eqn:1Dvector}) composed of obvious two-component parts
$\ba_\pm^T=(\balp^T_\pm,\bbet^T_\pm)$ we find that the expectation values $\balp^\dag_\pm\sigma_z\balp_\pm=0$ and
$\bbet^\dag_\pm\sigma_z\bbet_\pm=0$ and therefore the warping correction to the 1D step state dispersion $\delta E_{k_x\pm}=0$ vanishes. We conclude that in first order in the warping scale $\lambda$ the 1D step state Hamiltonian Eq.~(\ref{eqn:1DHam}) is unchanged, therefore we can expect that the possible appearance of Majorana end states in the SC case as discussed in the next section is also unaffected by the warping term in this order for all momenta $k=k_x$. Furthermore in the limit $k\rightarrow 0$ the warping perturbation effect vanishes intrinsically.

\section{Majorana zero modes in the SC proximity induced gap of 1D step states}
\label{sect:majorana}

In each of the two 1D bands of quasiparticles confined to the step the helical spin locking is protected since
it is inherited from the 2D topological surface states. Therefore they may be considered as spin-locked
1D excitations. If they  open a superconducting gap originating from the proximity effect of a superconducting substrate it is natural to expect \cite{alicea:12,beenakker:13,elliott:15,sato:16,chamon:10,marra:21,laubscher:21} the possible creation of Majorana zero modes (MZM)  at the end of the step line which would lead to a zero-bias conductance peak in transport and tunneling experiments \cite{das:12,jeon:17,aguado:20}. 
Due to the locking of opposite spins in the helical state an underlying conventional singlet or s-wave superconductor with gap $\Delta_0$ may be used instead of the difficult to realize p-wave superconductor necessary in spinless models \cite{alicea:12,cook:11,laubscher:21}. The topological state of a general 1D superconducting fermionic system is given by  a topological invariant \cite{kitaev:01,elliott:15} $\cal{M}$$=(-1)^\nu$ where $\nu$ is the number of Fermi points including band degeneracy in one half of the BZ for the normal state. For the considered model (Eq.~(\ref{eqn:E1D})) with one nondegenerate band we have $\nu=1$ and then $\cal{M}$$=-1$ characterizes a nontrivial topology of the superconducting state which may host Majorana zero modes as end states of the step.
They are described by zero-energy solutions of Bogoliubov-deGennes (BdG) equations
inside the SC gap and are characterised by quasiparticle operators which are identical to their conjugates.  Previous scenarios to create Majorana states have mostly involved TI wires \cite{cook:11,cook:12} on top of an s-wave SC and a flux passing through them to  obtain the constituent 1D excitations from which Majorana states are formed. The present proposal is comparatively simple, it just takes the stepped interface between an s-wave SC and TI with the step playing the role of the flux penetrated wire.
%
%

\subsection{Model for 1D BdG Hamiltonian of step states}

Due to the proximity effects the spin-singlet pairs of the substrate can propagate a certain distance into the normal state of the TI which is governed by its coherence length of the latter \cite{deGennes:66}. In the pure normal state it is given by $\xi_n=(\hbar v/2k_BT)$ which becomes large for low temperatures so that one may expect a still sizeable s-wave  gap $\Delta_0'< \Delta_0$ on the T,B surfaces of the topological insulator  thin film. In order to formulate the effective 
BCS Hamiltonian for the 1D step states, however, we have to be aware that the proximity effect works on all 2D TI surface states.
Therefore we first express the 1D Cooper pair operators formed from Eq.~(\ref{eqn:1Dquasi}) by the 2D operator basis. We must keep in mind that the two $(\la =\pm)$ 1D dispersions fulfill $E_{-k\la}=-E_{k\la}=E_{k-\la}$, therefore only {\it inter-band} pairing of states with opposite momenta and approximately equal energy are possible. For those pairs we have:
\be
\chi_{k+}\chi_{-k-}=\frac{i}{4}\sum_{\alpha=T,B}(c^\alpha_{k\ua}c^\alpha_{-k\da}-c^\alpha_{k\da}c^\alpha_{-k\ua})
=-\chi_{k-}\chi_{-k+}
\label{eqn:interpair}
\ee
This means the inter-band pairing of spin-locked 1D quasiparticles naturally results from the proximity  induced spin-singlet pairing of 2D surface states. Hereby we neglected on the right side  i) triplet terms with equal amplitude since they cannot be induced by the s-wave substrate and ii) inter-(T,B) surface pairing of 2D helical states which is difficult to justify on the basis of the proximity effect.
Then the pair amplitude is given by 
\be
\bl
\bra \chi_{k+}\chi_{-k-}\ket
=&
-\bra\chi_{k-}\chi_{-k+}\ket
\\=&
\frac{i}{4}\sum_{\alpha=T,B}\bra c^\alpha_{k\ua}c^\alpha_{-k\da}-c^\alpha_{k\da}c^\alpha_{-k\ua}\ket
\\
\sim& 
\frac{i}{4}(\Delta_T+\Delta_B)=i\Delta_0',
\el
\ee
where $\Delta_{T,B}$ are the s-wave order parameters on the two TI surfaces introduced by the proximity effect. If $\xi_n\gg d$ they may be almost equal. The proximity induced superconducting pair potential for the 1D surface states is then
\be
\bl
{\cal H}_{SC}
=
&\,
i\Delta_0'[\chi_{k+}\chi_{-k-}-\chi_{k-}\chi_{-k+}]
\\
&
-i\Delta_0^{'*}[\chi_{-k-}^\dag\chi_{k+}^\dag -\chi_{-k+}^\dag\chi_{k-}^\dag]
.
\el
\ee
Introducing the Nambu spinors for the two 1D bands according to $\bchi_k^{tr}=(\chi_{k+},\chi_{k-},\chi_{-k+}^\dag,\chi_{-k-}^\dag)$ and adding the normal quasiparticle part in Eq.~(\ref{eqn:1DHam}) we obtain for the total 1D step state BCS Hamiltonian
\be
\bl
&
{\cal H}=
{\cal H}_{ST}+{\cal H}_{SC}
=
\sum_k\bchi_k^\dag \tilde{h}_k\bchi_k;\;\;\;
\\
&
\tilde{h}_k=(vk - \mu)\la_z\tau_0 -\la_y( {\rm Re} \Delta'_0\tau_x+ {\rm Im} \Delta'_0\tau_y).
\el
\ee
Here the $\lambda$- and $\tau$ Pauli matrices act in the space of 1D bands $(\pm)$ and Nambu particle-hole space, respectively . For simplicity we first set $\mu=0$ i.e. the chemical potential lies at the Dirac point in Fig.~\ref{fig:gapping}, the case for general $\mu$ is treated at the end of this section. The explicit matrix form of $\tilde{h}_\bk$ is shown below. The step 1D quasiparticle energies in the superconducting state are then given by
\be
\tE_k=\pm[(vk)^2+|\Delta'_0|^2]^\fs,
\ee
where generally $|\De'_0|<|t_L|,|t_R|$, i.e. the SC gap is inside the larger 2D hybridisation gap, except when $d$ is  close to nodal points of $t_{L,R}(d)$ (Fig.~\ref{fig:hybridisation}a).

\subsection{MZM in-gap states at the step ends}

%
\begin{figure}[t]
\begin{center}
\includegraphics[width=0.98\columnwidth]{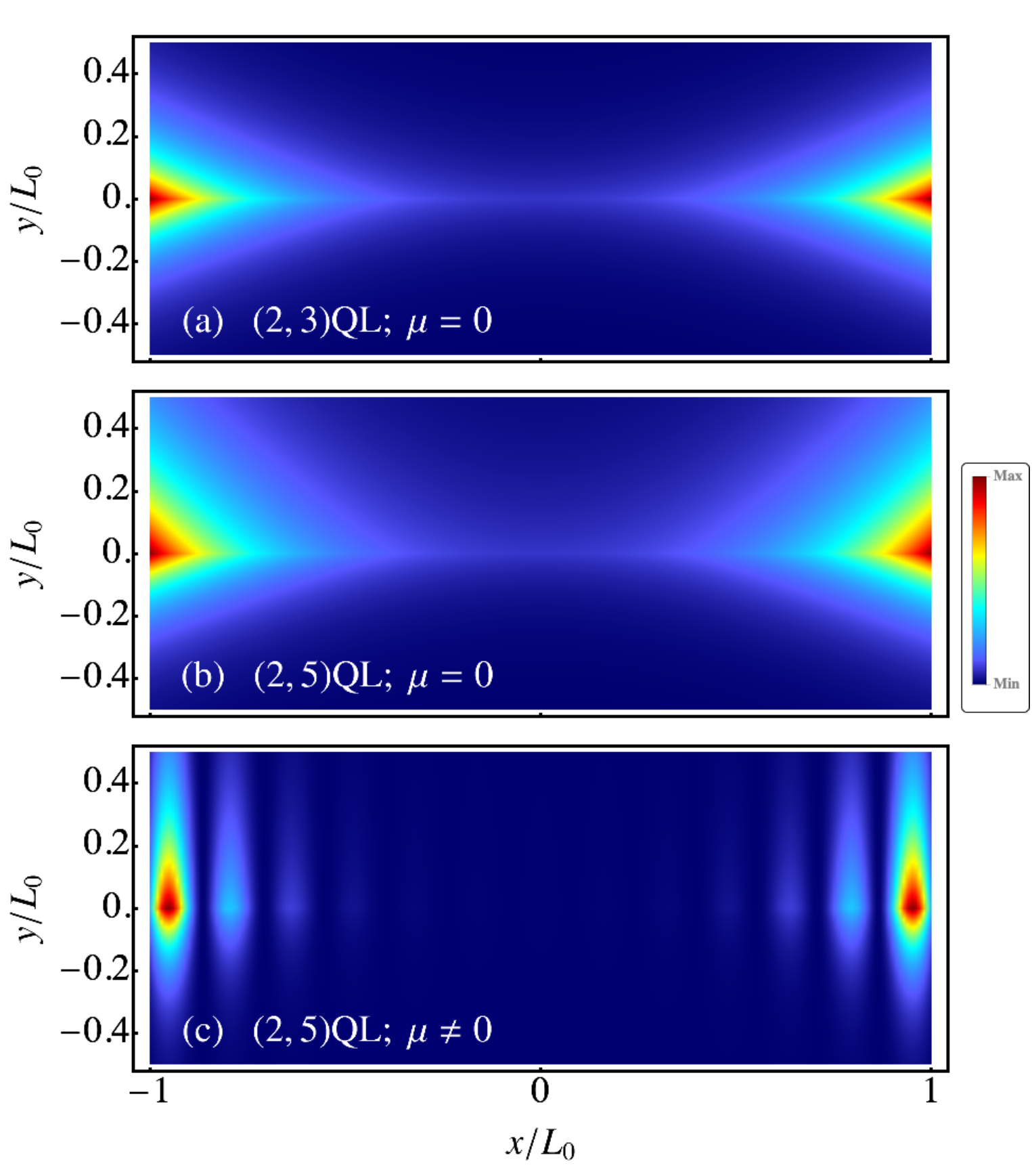}
\end{center}
\vspace{-0.5cm}
\caption{Density profile $p(x,y)$ of Majorana end states (superposition of u, d)) on the top (T) surface plane for the cases $(d_L,d_R)=(2,3); (2,5)$ for chemical potential $\mu$. Asymmetry and localization degree changes notably between the different pairs.
For nonzero $\mu$ MZM oscillations appear according to Eq.~(\ref{eqn:majosc}).
Here we set SC gap size $|\Delta_0'|=0.05 [E^*]$, $\mu= 0.25[E^*]$ in the lower panel and step length $2L_0=80 [\mbox{QL}]$. For presentation the latter is chosen artificially small.}
\label{fig:majorana}
\end{figure}
%

Now we will search for zero energy solutions of the BdG Hamiltonian inside the SC gap and whose wave function has to be located within the step length $-L_0\leq x \leq L_0$, where $\Delta'_0$ is finite.
Replacing $k\rightarrow -i\partial_x$ in the above Hamiltonian we obtain the equation
\be
\bigl[(v\la_z\tau_0\partial_x -i\la_y({\rm Re} \Delta'_0\tau_x+ {\rm Im} \Delta'_0\tau_y)\bigr]\tilde{\bphi}(x)=0
.
\label{eqn:BdGx}
\ee
In the space of four dimensional Nambu spinors  $\bchi_k$ the real space BdG Hamiltonian of Eq.~(\ref{eqn:BdGx}) is represented as the matrix 
\be
\bl
\tilde{h}(x)=
\left(
 \begin{array}{cccc}
-iv\partial_x& 0&0&i\De_0^{'*}\\
 0 & iv\partial_x&-i\De_0^{'*}&0\\
 0&i\De'_0&-iv\partial_x&0\\
 -i\De'_0&0& 0&iv\partial_x
 \end{array}
\right).
\el
\ee
Aside from a relative sign this matrix consists of two identical blocks due to the relation between inter-band pairings given in Eq.~(\ref{eqn:interpair}) and neglect of intra-band pairs.
Using this matrix form which factorises into two blocks the BdG equation (Eq.~(\ref{eqn:BdGx})) may be solved by the ansatz of Eqs.(\ref{eqn:xansatz},\ref{eqn:endstate}) assuming $\De'_0(x)=\De'_0\Theta(L_0+x)\Theta(L_0-x)$, i.e. constant $\De'_0$ within the sample. The two solutions for each block decay exponentially either along $x$ or $-x$. This means they will be
located at one of the ends of the step at $-L_0$ or $L_0$  which we call $d$(down) and $u$(up), respectively (Fig.\ref{fig:stepscheme}). For the two blocks $A,B$ we arrive at the zero energy BdG wave functions $(\nu=u,d)$
\be
\bl
\tilde{\bphi}_{A}^\nu(x)
=&\,
\ba^\nu_{A}w_\nu(x);
\;\;\;
\tilde{\bphi}_{B}^\nu(x)=\ba^\nu_{B}w_\nu(x);
\\
w_d(x)=
&\,
Ce^{\la(L_0+x)}; 
\;\;\;
 w_u(x)=Ce^{\la(L_0-x)},
\label{eqn:xansatz}
\el
\ee
where $C=[2\la/(1-exp(-4\la L_0))]^{-\fs}$. The decay length $\lambda^{-1}$ of the end states is given by $\la=\frac{|\Delta'_0|}{v}$ which may be written as $\la=\frac{1}{\pi}\frac{|\Delta'_0|}{\De_0}\frac{v^s_F}{v}\xi^{-1}$ which is proportional to the inverse BCS coherence length $\xi$ of the SC substrate. Here the first factor describes the gap reduction factor due to proximity effect and the second one the ratio of Fermi velocities in substrate $(v^s_F)$ and TI $(v)$. Furthermore the amplitude vectors of the wave functions are 
\be
\bl
(\ba^u_A)^{tr}&=\frac{1}{\sqrt{2}}(1,0,0,1);\;\;
(\ba^d_A)^{tr}=\frac{1}{i\sqrt{2}}(1,0,0,-1),
\\
(\ba^u_B)^{tr}&=\frac{1}{\sqrt{2}}(0,1,1,0);\;\;
(\ba^d_B)^{tr}=\frac{1}{i\sqrt{2}}(0,1,-1,0).
\label{eqn:endstate}
\el
\ee
The complete zero energy wave functions in the SC gap, including the $y$-dependence from Eq.~(\ref{eqn:stepstate}), is then
given by
\bea
\label{eqn:Mwave}
\tilde{\bphi}_{A,B}^\nu(x,y)&=&
\ba^\nu_{A,B}w_\nu(x,y)\\
w_\nu(x,y)&=&w_\nu(x)
\frac{1}{\nu_0}[e^{\kappa_Ly}\Theta(-y)+ e^{-\kappa_Ry}\Theta(y)],\non
\eea
where $-L_0\leq x,y\leq L_0$.
The wave functions for $A,B$ blocks are related by permutation of the two 1D band states according to $P\tilde{\bphi}^\nu_{A,B}=\tilde{\bphi}^\nu_{B,A}$ defined by the permutation operator $P=\lambda_x\tau_0$ which is a symmetry of the Hamiltonian.  Therefore the  wave functions of  the zero energy states should also be symmetrized by taking the combination $\tilde{\bphi}^\nu_{A,B}+P\tilde{\bphi}^\nu_{A,B}$ meaning
\be
\bl
\tilde{\bphi}_\nu(x,y) = \tilde{\bphi}^\nu_A(x,y)+\tilde{\bphi}^\nu_B(x,y)
\;\;\mbox{and}\;\;
\ba_\nu=\ba_A^\nu+\ba_B^\nu
.
\label{eqn:Msymm}
\el
\ee
For these symmetrised zero energy wave functions we may construct their corresponding field operators according to
\be
\bl
\Psi_u
&=\!
\int dx\tilde{\bphi}^\dag_u(x)\bchi_x
=\!
 \int dxw_u(x)\ba_u^\dag\bchi_x
 =\!
 \int dx w_u(x)\gamma_1,
\\
\Psi_d
&=\!
\int dx\tilde{\bphi}^\dag_d(x)\bchi_x
=\!
 \int dxw_d(x)\ba_d^\dag\bchi_x
 =\!
 \int dx w_d(x)\gamma_2,
\label{eqn:Mop}
\el
\ee
where we used the real space representation $\bchi_x=\int dke^{-ikx}\bchi_k$. These operators satisfy the
reality condition $\Psi_u^\dag=\Psi_u$ and  $\Psi_d^\dag=\Psi_d$ and 
therefore represent two Majorana zero modes separated at the two ends $(d,u)$  of the step length.
Their associated quasiparticle operators $\gamma_1$, $\gamma_2$ corresponding to zero energy states 
with BdG wave functions  $\tilde{\phi}_\nu(x,y)$ in Eqs.~(\ref{eqn:Mwave},\ref{eqn:Msymm}) are then obtained as
\be
\bl
\gamma_1
&=
\ba_u^\dag\bchi_x=\frac{1}{\sqrt{2}}(\chi_{x+}+\chi_{x-}+\chi_{x+}^\dag+\chi_{x-}^\dag);
\\
\gamma_2
&=
\ba_d^\dag\bchi_x=\frac{i}{\sqrt{2}}(\chi_{x+}+\chi_{x-}-\chi_{x+}^\dag-\chi_{x-}^\dag),
\label{eqn:majo1}
\el
\ee
which fulfil the reality condition $\gamma_i^\dag=\gamma_i \;\;(i=1,2)$. These quasiparticle operators for the MZM  satisfy the canonical Majorana anti-commutation rules $\{\gamma_{1},\gamma_{1}\}=\{\gamma_{2},\gamma_{2}\}=2$ and $\{\gamma_{1},\gamma_{2}\}=0$.

The real-space density profile $p(x,y)=|w_u(x,y)+w_d(x,y)|^2$
of these SC in-gap zero- energy states is governed by three (inverse) length scales: i) perpendicular to the step by $\kappa_L(d_L)=sign(t_L)t_L(d_L)/v$ and    $\kappa_L(d_R)=sign(t_R)t_R(d_R)/v$  to the right and left and ii) parallel to the step by $\lambda=|\Delta'_0|/v$. This means that the Majorana density profile can be quite anisotropic and change rapidly as function of the relative
TI film thicknesses $d_L$, $d_R$ on both sides of the step. In the general case when both $t_L(d_L)$ and $t_R(d_R)$ are not located close to the zeroes of the oscillatory function in Eq.(\ref{eqn:tosc}) one has $|\Delta'_0| < |t_L|, |t_R|$ and therefore the Majorana profile is concentrated at the step with a more gradual decay along step direction $x$ (Fig.~\ref{fig:majorana}).\\

Finally we briefly give the results for the case of general position of the chemical potential $\mu\neq 0$ inside the hybridization gap (Fig.~\ref{fig:gapping}(c)), cutting the 1D dispersions at finite wave vector $k_c=\mu/v$ once in the positive half of the BZ.  In the SC state this leads to quasiparticle bands
\bea
\bl
\tE^\pm_{k1,2}=\pm[(vk\pm\mu)^2+|\Delta'_0|^2]^\fs,
\el
\eea
which have now two branches $(1,2)$ emerging from $k_{1,2}=\pm k_c$ and a SC gap $2|\Delta'_0|$, independent of the position of the chemical potential inside the hybridisation gap. The solutions for the MZM end states may be found in analogy to the above derivation. The essential difference lies in the MZM wave functions now given by 
\bea
\bl
w_u(x)=&Ce^{\lambda_1(x-L_0)}e^{i\lambda_2x};
\\
w_d(x)=&Ce^{-\lambda_1(x+L_0)}e^{i\lambda_2x},
\label{eqn:majosc}
\el
\eea
where $\lambda_1=|\Delta'_0|/v$ and $\lambda_2=\mu/v$. Thus, in addition to the exponential decay from
the ends there is an oscillation of MZM form factors which becomes faster than the decay for $|\mu|>|\Delta'_0|$.
Analogous to Eq.~(\ref{eqn:Mop}) the self-adjoint field operators of MZM modes can be written as
\bea
\bl
\Psi_u=&\frac{1}{2L_0}\int dx[w'_u(x)\gamma_1+w''_u\tilde{\gamma_1}];
\\
\Psi_d=&\frac{1}{2L_0}\int dx[w'_d(x)\gamma_2+w''_d\tilde{\gamma_2}],
\el
\eea
where we split $w_{u,d}(x)=w'_{u,d}(x)+iw''_{u,d}(x)$ into real and imaginary parts. we have now an additional Majorana pair $(\tilde{\gamma}_1,\tilde{\gamma_2})$ complementing Eq.~(\ref{eqn:majo1}) and defined by the quasiparticle operators 
\be
\bl
\tilde{\gamma_1}
&=
\frac{1}{i\sqrt{2}}(\chi_{x+}-\chi_{x-}-\chi_{x+}^\dag+\chi_{x-}^\dag);
\\
\tilde{\gamma_2}
&=
\frac{1}{\sqrt{2}}(\chi_{x+}-\chi_{x-}+\chi_{x+}^\dag-\chi_{x-}^\dag),
\el
\ee
The doubling  is related to the existence of two Bogoliubov excitation bands for $\mu\neq 0$ arising from from $k_{1,2}=\pm k_c$ points.  Again the Majorana anti-commutation rules $\{\tga_{1},\tga_{1}\}=\{\tga_{2},\tga_{2}\}=2$ and $\{\tga_{1},\tga_{2}\}=0$ are fulfilled, furthermore we have  $\{\gamma_{1},\tga_{1}\} =  \{\gamma_{2},\tga_{2}\} =0$. The density profiles are $p(x,y) =|w'_u(x,y)+w'_d(x,y)|^2$ and $\tilde{p}(x,y)=|w''_u(x,y)+w''_d(x,y)|^2$ with $w_\nu(x,y)=w_\nu(x)\frac{1}{\nu_0}[e^{\kappa_Ly}\Theta(-y)+ e^{-\kappa_Ry}\Theta(y)]$ similar as in Eq.~(\ref{eqn:Mwave}). An example of the MZM profile for nonzero $\mu$ is shown in the bottom of Fig.~\ref{fig:majorana} which exhibits the additional oscillatory behaviour.

\section{Conclusion and outlook}

In this work we have shown that the helical surface states in thin films of topological insulators may be manipulated in an
interesting and promising way. It has previously been theoretically derived on general grounds that in thin films the surface
states exhibit hybridisation due to inter-surface wave function overlap. This leads to a hybridisation energy that both decays
exponentially and oscillates with film thickness depending on materials parameters and simultaneously leads to a gapping
of the 2D helical surface states.

This can be exploited in a simple
way by profiling the film thickness in a suitable manner, for example by introducing a step in film thickness via the substrate 
such that the inter-surface hybridisation has opposite sign on both sides of the step. This leads to the appearance
of novel type of  non-degenerate 1D helical states confined spatially to the step with linear dispersion and again helical spin locking.
These states are protected by the sign change of the hybridisation and their decay perpendicular to the step is controlled by
the modulus of the hybridisation energy. It should be possible to investigate the existence of these 1D step states and their dispersion
by STM spectroscopy. Before this, however, it would be useful to check by STM and ARPES  experiments  whether homogeneous thin films indeed exhibit the theoretically predicted and prerequisite oscillations in hybridisation energy. This should immediately translate in the (rectified) oscillations of the gap size of coupled surface states. While \bse~does not seem to exhibit such oscillations the theoretical band parameters for \bte~and \ste~are apparently more favourable for their appearance.

If these conjectures can be experimentally  verified in some cases another highly attractive possibility opens up: When the substrate becomes a simple s-wave superconductor the proximity effect leads to induced gapping of the 1D helical step states. Since the latter are nondegenerate fermions with spin locking this can create Majorana type zero-energy modes inside the proximity effect induced gap which are localised at the end of the steps where the gap drops to zero. Their inverse localisation lengths along and perpendicular to the step is proportional to the SC gap energy and left/right inter-surface hybridisation energies. These Majorana end states should, like 1D step states themselves be observable with STM spectroscopy. The present proposed scenario for creating Majorana states is exceedingly
simple, requiring only a suitably profiled (s-wave superconducting substrate) to create the 1D step states. Since the latter are
already nondegenerate by their helical spin-locked nature one should not need additional arrangements like applied magnetic fields
or applying additional ferromagnetic layers which have been discussed before in the wire-type geometries for creating Majorana states.

The present simplistic geometry may be replaced by more elaborate ones. For example the step may not be extended to the whole width $[-L_0,L_0]$  of the sample but may only be present for $|x|\leq x_0 <L_0$. When $|x|$ approaches $x_0$ the thicknesses $d_R,d_L$ may be gradually changed on both sides of the step to a common $d_0$  which satisfies $t_R(d_0)=t_L(d_0)=0$. Then the 1D helical step state will also cease to exist at $\pm x_0$ position within the thin film area. This would presumably simplify investigations by STM method and suppress unwanted effects from sample ends. Another extension would be a regular array of steps that are a certain distance $y_0$ apart leading to sign change of $t(y)$ with period $2y_0$.
This configuration would be suitable for studying the effects of 1D step state overlap. Finally one might form a ring-like step, i.e. a quantum dot with the proper thickness inside and outside the ring to support a 1D ring state. 
This should lead to a discretization of the linear dispersion of step states, depending on the diameter of the ring and the thickness variation characteristics.
It is therefore certainly worthwhile to study these configurations and the possible in-gap (hybridisation, superconducting) states and their physical consequences further.


\section*{Acknowledgments}
  A.~A. acknowledges the support
 of the Max Planck- POSTECH-Hsinchu Center for Complex Phase
 Materials, and financial support from the National Research
 Foundation (NRF) funded by the Ministry of Science of Korea (Grant
 No. 2016K1A4A01922028). 


\appendix



\bibliography{References}

\end{document}